\documentclass{article}

\usepackage{PRIMEarxiv}
\usepackage[numbers]{natbib}
\usepackage{amsmath}
\usepackage[utf8]{inputenc} 
\usepackage[T1]{fontenc}    
\usepackage{hyperref}       
\usepackage{url}            
\usepackage{booktabs}       
\usepackage{amsfonts}       
\usepackage{nicefrac}       
\usepackage{microtype}      
\usepackage{lipsum}
\usepackage{fancyhdr}       
\usepackage{graphicx}       
\graphicspath{{media/}}     

\title{Resilient Biosecurity in the Era of AI-Enabled Bioweapons

}

\author{
  Jonathan Feldman \\
  Georgia Institute of Technology \\
  Atlanta, Georgia\\
  \texttt{jonathanfeldman@gatech.edu} \\
   \And
  Tal Feldman \\
  Yale Law School \\
  New Haven, CT\\
  \texttt{tal.feldman@yale.edu} \\
}

\begin{document}
\maketitle

\begin{abstract}
Recent advances in generative biology have enabled the design of novel proteins, creating significant opportunities for drug discovery while also introducing new risks, including the potential development of synthetic bioweapons. Existing biosafety measures primarily rely on inference-time filters such as sequence alignment and protein–protein interaction (PPI) prediction to detect dangerous outputs. In this study, we evaluate the performance of three leading PPI prediction tools: AlphaFold 3, AF3Complex, and SpatialPPIv2. These models were tested on well-characterized viral–host interactions, such as those involving Hepatitis B and SARS-CoV-2. Despite being trained on many of the same viruses, the models fail to detect a substantial number of known interactions. Strikingly, none of the tools successfully identify any of the four experimentally validated SARS-CoV-2 mutants with confirmed binding. These findings suggest that current predictive filters are inadequate for reliably flagging even known biological threats and are even more unlikely to detect novel ones. We argue for a shift toward response-oriented infrastructure, including rapid experimental validation, adaptable biomanufacturing, and regulatory frameworks capable of operating at the speed of AI-driven developments.\end{abstract}


\section{Introduction}
\noindent In recent years, the convergence of generative AI and molecular biology has produced powerful new systems capable of designing proteins with remarkable precision. Protein language models (PLMs), along with structural predictors and design tools like RFdiffusion, are revolutionizing medicine by enabling rapid therapeutic development—from novel enzymes and vaccines to antibody candidates tailored for specific diseases \cite{ruffolo_designing_2024,CHEN2023706,xiao2025proteinlargelanguagemodels,NEURIPS2021_f51338d7,hie_evolutionary_2022}. These systems are accelerating biomedical innovation, shrinking experimental timelines, and expanding the reach of drug design to previously intractable targets \cite{watson_novo_2023,bielska2025applyingcomputationalproteindesign}.

Yet this same capability introduces an unprecedented dual-use dilemma. The tools now used to craft life-saving biologics can, in the wrong hands, be repurposed to generate pathogens with enhanced infectivity, immune evasion, or toxicity \cite{dianzhuo_wang_without_2025}. PLMs can simulate the evolutionary trajectory of viral proteins, suggesting mutations that make them more transmissible or harder to neutralize \cite{huot_few-shot_2025,youssef_computationally_2025,cheng_zero-shot_2024}. Structure-aware design tools can validate these outputs in silico and propose new scaffolds to support them \cite{watson_novo_2023,hayes_simulating_2024}. This design-and-validation loop—which mirrors workflows already used in therapeutic biotech—lowers the barrier to creating synthetic threats, some of which may be novel enough to evade detection altogether \cite{dianzhuo_wang_without_2025}.

In response to this growing concern, the AI and biosecurity communities have proposed technical countermeasures \cite{wang_call_2025}. These include inference-time filtering to block dangerous outputs, red-teaming to test models for misuse, and sequence screening to flag potentially pathogenic molecules \cite{dianzhuo_wang_without_2025,wang_call_2025,doi:10.1126/science.ado1671}. However, these methods rest on a problematic assumption: that we can reliably identify harmful proteins computationally.

This assumption breaks down under scrutiny. Whether a protein is benign or harmful depends on factors that remain opaque to even the most advanced models—its tertiary structure, cellular context, binding specificity, and immunogenicity \cite{dianzhuo_wang_without_2025,doi:10.1126/science.ado1671,jeffery_current_2023}. Most of these properties cannot be inferred from sequence alone. In fact, as we demonstrate in this paper’s technical section, existing models for protein–protein interaction (PPI) prediction fail especially badly when applied to viral proteins. They lack the data diversity, contextual understanding, and biological realism to generalize beyond narrow, overrepresented domains. This limitation creates blind spots that could be exploited by malicious actors or simply overlooked in good-faith deployments.

If computational filters are not fully reliable then only experiments can provide definitive safety assurance. Yet experimental validation methods, such as binding assays, toxicity screens, and infection models, are too slow, costly, and resource-intensive to scale with model outputs \cite{fowler_deep_2014,ueki_antibody_2024}. In theory, flagged sequences could be tested in the lab. In practice, we lack the infrastructure to vet more than a tiny fraction \cite{dianzhuo_wang_without_2025,zhu_toxdl_2025}. That leaves us with a dangerous asymmetry: generative models can produce threats at scale, but our ability to verify them lags far behind.


Worse, even dramatic advances in our ability to predict pathogenicity from sequence data will not guarantee security. Open-source models, fine-tuned on public data, are available worldwide. The barriers to misuse are falling: DNA synthesis services are increasingly accessible, foundational laboratory tools are widely distributed, and the necessary expertise is no longer confined to state-sponsored programs \cite{dianzhuo_wang_without_2025,hayes_simulating_2024,passaro_towards_nodate,kosuri_large-scale_2014,10.1093/nar/gkw1027}. Nation-states and non-state actors with sufficient resources can run models privately, circumvent safety mechanisms, and synthesize outputs with little oversight. In this threat landscape, technical success in sequence classification does not guarantee security. We cannot rely on preemptive filters alone.

The implication is stark: current defense strategies—focused on prediction, prevention, and perimeter control—are structurally mismatched to the nature of the threat. No matter how refined our red-teaming becomes, no matter how many inference-time filters we deploy, we cannot guarantee safety when our core tools lack ground truth and when potential adversaries are unconstrained.

What is needed instead is a shift: from a paradigm of containment to one of resilience. Rather than attempting to intercept every potentially dangerous output at generation time, we must build systems capable of responding to threats after they emerge—faster than those threats can spread. This means developing AI-native infrastructure for therapeutic response: models trained to design safe, stable, and manufacturable countermeasures on demand; high-throughput simulation and validation tools that can screen thousands of candidates in parallel; and a nationwide biomanufacturing pipeline that can pivot from peacetime research to emergency deployment within days.

Crucially, this is not merely a technical aspiration—it is a national security imperative. We must prepare for a world in which synthetic pathogens may be generated and released faster than traditional public health systems can detect or contain them. In such a world, resilience is the only viable defense. Speed—not restraint—becomes the battlefield.

This paper is structured as follows: In the next section, we show that current machine learning models for predicting viral protein interactions, especially those relying on large-scale sequence and structure modeling, fail to generalize well. This is particularly evident in low-data and high-divergence scenarios. These shortcomings highlight the limitations of current safety frameworks and underscore the urgent need to shift research focus towards response-driven capabilities. We then propose a resilience-based framework that addresses gaps in existing biosecurity efforts. Ultimately, our goal is not to eliminate generative biology, but to guide it—ensuring that the tools being developed today are aligned with the future needs that will arise.

\section{Limitations of Current Inference-Time Filtering Approaches}

The task of identifying potentially pathogenic proteins, especially those that may be synthetically engineered or genetically modified, for inference-time filtering requires robust, reliable tools. In the context of biosecurity, these tools must be able to predict whether a given protein sequence or structure could be harmful to humans. Currently, the two primary methodologies used to predict pathogenicity are sequence similarity comparisons and PPI prediction mechanisms \cite{doi:10.1126/science.ado1671,gonzales_phistruct_2025,hu_spatialppiv2_2025,doi:10.1128/jvi.02595-13}. While these tools have proven useful, both are fundamentally limited in their applicability and accuracy, especially when it comes to detecting novel or engineered proteins.

Sequence similarity searches are arguably the most straightforward and effective tool for identifying known pathogenic proteins. By comparing an input protein sequence against a large database of known sequences, researchers can rapidly identify homologous proteins from established viral strains \cite{Steinegger079681}. This method is particularly useful for identifying pathogenic proteins that have undergone minor mutations or are derivations of previously known viral proteins. Tools such as BLAST facilitate the comparison of protein sequences to databases such as UniProt and NCBI Virus, which contain thousands of viral sequences \cite{ALTSCHUL1990403,10.1093/nar/gku1207,10.1093/nar/gkp846}. These comparisons allow for the determination of the degree of homology between the input protein and known viral sequences, providing a relatively simple and highly scalable means of detecting pathogenicity \cite{ALTSCHUL1990403,mirdita_colabfold_2022}. When the degree of homology exceeds a pre-defined threshold, the protein in question can be flagged as potentially pathogenic.

Despite its effectiveness, sequence similarity searches are fundamentally limited by their reliance on existing databases. Since these databases are composed of known sequences, they inherently fail to detect novel viral proteins that are sufficiently divergent from anything previously cataloged \cite{doi:10.1126/science.ado1671}. This means that while sequence similarity searches can effectively identify variants of known pathogens, they are incapable of detecting entirely novel proteins, such as those designed using advanced computational techniques like RFdiffusion or ESM3, which generate novel protein sequences that are not present in any existing databases \cite{watson_novo_2023,hayes_simulating_2024}. Therefore, while sequence similarity searches remain a valuable tool for detecting known pathogens, they cannot be relied upon to identify entirely new threats, especially those that have been deliberately engineered.

To address this limitation, researchers have turned to PPI prediction methodologies \cite{hashemifar_predicting_2018,nikam_deep_2023}. PPIs are central to the pathogenicity of many viruses, as viral proteins typically need to interact with human cellular receptors to gain entry into the host \cite{gao_improved_2024,thadani_learning_2023,MAGINNIS20182590}. While most viruses bind to standard receptors, some also interact with glycoproteins and glycolipids on the surface of host cells \cite{MAGINNIS20182590}. PPI prediction tools have been developed with the aim of identifying these interactions, thereby providing a complementary approach to sequence similarity searches. In theory, PPI prediction tools could be used to determine whether a viral protein is likely to bind to a receptor on a human cell, serving as a reasonable proxy for pathogenicity. This would allow for the identification of proteins that might be capable of interacting with human receptors, even if they are not closely related to any known pathogen.

However, in practice, PPI prediction tools—especially those based on artificial intelligence—often fail to deliver the accuracy required for high-stakes applications such as biosecurity \cite{10.1093/bioinformatics/btae012}. The inherent limitations of these models are particularly concerning when one considers the potential consequences of inaccuracies. A false negative, in which a dangerous pathogenic protein goes undetected, could have dire consequences for public health. For these reasons, the accuracy of PPI prediction models is a critical issue that must be addressed before these tools can be used reliably for biosecurity purposes.

In order to evaluate the performance of state-of-the-art PPI prediction models, we conducted a benchmarking study using thirteen well-characterized viral protein–human receptor interactions. These interactions were selected from viruses that have fully or partially resolved structures available in the Protein Data Bank (PDB), such as HIV, Measles, and SARS-CoV-2 \cite{10.1093/nar/gky1004}. Three widely used PPI prediction tools were selected for testing: AlphaFold 3, AF3Complex, and SpatialPPIv2. AlphaFold 3, developed by Google DeepMind, is a deep learning-based model that is widely regarded as the best available tool for predicting protein structures \cite{abramson_accurate_2024}. AF3Complex is a derivative of AlphaFold 3 that has been specifically designed to improve PPI predictions by modeling those interactions more accurately \cite{feldman_af3complex_2025}. SpatialPPIv2 is a graph neural network-based model that represents the latest advancement in PPI prediction methodologies \cite{hu_spatialppiv2_2025}.

Each of the models was provided with the protein sequences of the viral binding protein and its corresponding human receptor. AlphaFold 3 and AF3Complex used sequence-based inputs to predict the structures of the individual proteins, while SpatialPPIv2 was provided with pre-computed structures generated by AlphaFold 3 \cite{hu_spatialppiv2_2025,abramson_accurate_2024,feldman_af3complex_2025}.SpatialPPIv2 differs from AlphaFold 3 and AF3Complex in that it is explicitly trained to predict PPIs using structural inputs rather than sequence information \cite{hu_spatialppiv2_2025}. AlphaFold 3 relies on its internally generated interface predicted TM-score (ipTM) to evaluate the confidence of predicted protein complex structures; prior studies have used an ipTM score of 0.6 or higher as a proxy for a meaningful PPI \cite{abramson_accurate_2024,wee_benchmarking_2024}. In contrast, AF3Complex uses a separate internal metric, the predicted Interface-Similarity score (pIS), with scores of 0.38 or above indicating a likely PPI \cite{feldman_af3complex_2025,Zhao2025.05.07.652715,gao_af2complex_2022}. SpatialPPIv2 instead outputs a probability that a meaningful interaction exists between the input proteins, with a threshold of 0.5 or higher used to indicate a predicted PPI.

\begin{table*}[ht]
\centering
\renewcommand{\arraystretch}{1.6} 
\setlength{\tabcolsep}{18pt} 
\begin{tabular}{|l|c|c|c|}
\hline
\textbf{Virus} & \textbf{SpatialPPIv2} & \textbf{AlphaFold3} & \textbf{AF3Complex} \\
\hline
Adenovirus & $\checkmark$ & $\checkmark$ & $\checkmark$ \\
\hline
Chikungunya & $\times$ & $\times$ & $\times$ \\
\hline
Ebola & $\checkmark$ & $\times$ & $\checkmark$ \\
\hline
Epstein-Barr & $\checkmark$ & $\checkmark$ & $\checkmark$ \\
\hline
Hepatitis B & $\times$ & $\times$ & $\times$ \\
\hline
HIV & $\checkmark$ & $\checkmark$ & $\checkmark$ \\
\hline
HSV (HVEM Receptor) & $\checkmark$ & $\times$ & $\times$ \\
\hline
HSV (Nectin-1 Receptor) & $\checkmark$ & $\checkmark$ & $\checkmark$ \\
\hline
Measles & $\times$ & $\times$ & $\times$ \\
\hline
Measles (Edmonston strain) & $\times$ & $\checkmark$ & $\checkmark$ \\
\hline
MERS & $\checkmark$ & $\checkmark$ & $\checkmark$ \\
\hline
Nipah & $\checkmark$ & $\checkmark$ & $\checkmark$ \\
\hline
SARS-CoV-2 & $\times$ & $\times$ & $\checkmark$ \\
\hline
\end{tabular}
\caption{Virus-host protein-protein interaction prediction performance across three state-of-the-art computational models. Each row represents a different virus strain or receptor variant, with checkmarks ($\checkmark$) indicating successful PPI identification and crosses ($\times$) indicating failed predictions. The models have different accepted confidence thresholds: SpatialPPIv2 (0.5), AlphaFold3 (0.6), and AF3Complex (0.38).}
\label{tab:virus_success}
\end{table*}

The results of the benchmarking study, summarized in Table~\ref{tab:virus_success}, reveal a significant failure of all three PPI prediction models to consistently identify well-documented viral–host interactions. Notably, both AlphaFold 3 and SpatialPPIv2 failed to correctly identify the interaction between the SARS-CoV-2 spike protein and the human ACE2 receptor—an interaction that is critical to the virus’s ability to infect human cells and that was at the heart of the COVID-19 pandemic \cite{huot_few-shot_2025,wang_deep-learning-enabled_2023}. In total, SpatialPPIv2 misidentified approximately 40\% of the PPIs, while AlphaFold 3 misidentified about 50\% and AF3Complex misidentified about 30\%. These results indicate that even state-of-the-art models are far from reliable when it comes to predicting viral–host interactions, which undermines the utility of these models for identifying computationally designed pathogenic proteins.
Moreover, many of the PPIs included in the benchmarking study were present in the training datasets of all three models \cite{abramson_accurate_2024,feldman_af3complex_2025,Kovtun2024.07.17.603980}. This means that these models had already been exposed to these interactions during training, yet they still failed to accurately predict them in the testing phase. This is a particularly troubling finding, as it suggests that the models are not generalizing well to real-world PPI predictions, even when the interactions are well-known and well-characterized. If these models cannot even accurately predict the most established viral PPIs, the possibility of them detecting novel or engineered viral–host interactions becomes even more unlikely.

As an ancillary experiment, we evaluated four mutated variants of the SARS-CoV-2 spike receptor-binding domain, the viral protein responsible for mediating COVID-19 infections. These mutations were sourced from a prior study \cite{moulana_compensatory_2022}, which experimentally quantified their effects on binding affinity between the spike RBD and the human ACE2 receptor, a key determinant of protein–protein interaction strength. Among the selected variants, two exhibited the greatest enhancement in ACE2 binding affinity, with more than tenfold increases, while the remaining two showed the most substantial reductions in affinity that still preserved a functional interaction.

We assessed these four spike mutants using the same three structure-based models evaluated previously—SpatialPPIv2, AlphaFold3, and AF3Complex—to determine whether any PPIs between spike and ACE2 were computationally predicted. The results are shown in Table~\ref{tab:affinity_model_scores}.

Strikingly, all three models failed to detect meaningful interactions for any of the variants, including those with experimentally confirmed high-affinity binding. This reveals a strong insensitivity to combinatorial mutational effects, highlighting a significant blind spot: in the absence of sequence similarity searches, these models may be unable to identify pathogenic viral variants. This raises serious concerns about their utility in validating viral pathogenicity.

Even more concerning is the direction of the models' confidence shifts. For both beneficial and detrimental mutations, SpatialPPIv2 and AlphaFold3 increased their predictive confidence, while AF3Complex decreased its own, regardless of the true biological impact. This erratic behavior underscores the models' limitations when faced with mutationally altered viral proteins—a critical vector through which more virulent or evasive pathogens could emerge \cite{dianzhuo_wang_without_2025}.

A major contributor to these failures is a fundamental data deficiency, a lack of high-quality, well-annotated examples of viral–host protein interactions in existing training datasets \cite{YUAN20252252,nomburg_birth_2024}. Although resources like the PDB contain hundreds of thousands of protein structures, the vast majority of these represent non-viral proteins, with viral proteins, especially those relevant to host interactions, comprising only a small fraction \cite{10.1093/nar/gky1004,nomburg_birth_2024}. This lack of targeted data limits the models' ability to learn interaction-relevant features specific to viral proteins. 

However, this data imbalance is only part of the challenge. Generalizing to novel or engineered proteins remains inherently difficult because of the complex biophysical nature of protein–protein interactions \cite{YUAN20252252}. Even with improved viral datasets, predicting interactions involving synthetic or previously unseen proteins, such as those generated by modern protein design tools, presents a major obstacle. These artificial proteins may not resemble known viral proteins and may interact with host receptors in unconventional ways, making them especially difficult for models trained on natural proteins to detect \cite{watson_novo_2023,hayes_simulating_2024}.

\begin{table*}[ht]
\centering
\renewcommand{\arraystretch}{1.8} 
\setlength{\tabcolsep}{20pt} 
\begin{tabular}{|c|c|c|c|}
\hline
\textbf{Affinity Change} ($\Delta \log_{10} K_D$)
& \textbf{SpatialPPIv2} 
& \textbf{AlphaFold3} 
& \textbf{AF3Complex} \\
\hline
$-1.04$ 
& 0.27 (+0.08) 
& 0.16 (+0.03) 
& 0.11 (--0.47) \\
\hline
$+2.14$ 
& 0.19 (+0.00) 
& 0.24 (+0.11) 
& 0.06 (--0.52) \\
\hline
$-1.03$   
& 0.26 (+0.07) 
& 0.14 (+0.01) 
& 0.06 (--0.52) \\
\hline
$+2.11$ 
& 0.24 (+0.05) 
& 0.15 (+0.02) 
& 0.05 (--0.53) \\
\hline
\end{tabular}
\caption{Performance comparison of PPI prediction models on Sars-Cov-2 spike protein mutants. Each row represents a different mutational variant with its corresponding change in binding affinity ($\Delta \log_{10} K_D$), with a negative value indicating an increase in binding affinity and a positive value indicating the opposite. The values in parentheses show the difference from the baseline confidence score for each model for the wildtype Sars-Cov-2 spike protein and human ACE2 PPI. SpatialPPIv2's threshold for PPI identification is 0.5, while those of AlphaFold 3 and AF3Complex are 0.6 and 0.38, respectively. \textit{No model of the three correctly predicted a single PPI.}}
\label{tab:affinity_model_scores}
\end{table*}

This gap underscores a broader limitation in current PPI prediction approaches. While models can perform reasonably well when applied to familiar protein types, they falter in high-stakes scenarios where generalization is critical, such as detecting engineered proteins with potential for harm. As generative tools become more accessible and capable, the inability to reliably assess the risk of novel proteins presents a serious and growing biosecurity concern.

\section{Reorienting Research Toward Infrastructure Readiness and  Rapid Response }

Current inference-time filters and protein–protein interaction predictors aren’t enough to catch dangerous synthetic proteins. As generative biological models improve and spread, it is increasingly clear that biosecurity cannot rely on static safeguards alone. Instead, the focus must shift toward building a resilient infrastructure that is capable of designing, validating, manufacturing, and deploying medical countermeasures at machine timescales. That means leadership from federal agencies but also a broad realignment of research priorities across universities, research labs, and industry.

To date, most investment in generative AI for therapeutic design has been concentrated in the pharmaceutical industry and biotechnology startups \cite{HUANG2023100446,reardon_five_2024}. These efforts naturally focus on common diseases with strong commercial incentives and well-known biological targets. But that leaves a gap: there’s little work on the rare, fast-evolving, or engineered pathogens most relevant to biodefense \cite{dianzhuo_wang_without_2025,de_lima_artificial_2024,de_haro_biosecurity_2024}. Academic research must fill this gap. Universities and consortia are well suited to study the edge cases that industry R\&D tends to overlook. By targeting novel viruses, unknown protein structures, and synthetic sequences outside today’s biological data, they can help build tools that matter for alignment and security.

These efforts will fail without addressing a deeper problem: the critical data needed to train and evaluate countermeasure models is scattered and siloed. Information on pathogen evolution, immune response, failed therapeutics, and synthetic constructs sits across disconnected academic, corporate, and government systems \cite{hummer_investigating_2025}. Integrating that data is essential—but so is access control. In a national security context, broad sharing is risky. Access should be limited to certified research labs and trusted partners operating under clear safeguards. We need sustained investment in shared infrastructure to support this: repositories of known and hypothetical viral strains, standardized measurements of immunogenicity and toxicity, and transparent clinical performance data. Funding agencies should prioritize these efforts, and academic institutions must be willing to build and maintain secure platforms that enable collaboration without compromising biosecurity. The COVID-19 response showed what rapid, coordinated data sharing can unlock \cite{doi:10.1177/17456916231178719}; biodefense now requires a version built for both speed and control.

Beyond data and modeling, the academic community has a responsibility to reframe how generative protein design is conceptualized. Much of the discourse surrounding PLMs and other generative tools has focused on speculative risks, particularly their potential misuse for creating harmful biological agents \cite{dianzhuo_wang_without_2025,wang_call_2025,de_lima_artificial_2024,de_haro_biosecurity_2024}. But the risks posed by generative protein models are no longer theoretical, and the research community needs to treat them accordingly. As these systems grow more capable, especially in designing novel or engineered proteins, their potential for misuse becomes a concrete concern. Addressing this requires more than superficial safeguards or ad hoc filtering. It demands structural solutions. Treating biosecurity as a secondary or post hoc issue is no longer tenable. These challenges sit at the center of responsible model development and must be integrated into core research agendas.

Still, while academia can help define the technical foundations of preparedness, the responsibility for real-world deployment rests firmly with the federal government. Wet-lab validation, clinical testing, and large-scale manufacturing of AI-generated therapeutics will require coordination, supply chains resilience, and emergency infrastructure that are beyond the capacity of any individual lab or institution. Agencies such as the Biomedical Advanced Research and Development Authority (BARDA), the Food and Drug Administration (FDA), and the Administration for Strategic Preparedness and Response (ASPR) must take the lead in establishing national infrastructure that translates model output into deployable countermeasures.

This includes building experimental screening platforms that can test thousands of protein designs in parallel for binding, toxicity, manufacturability, and immune response. These systems should be fast, modular, and automated to keep pace with rapid model outputs. But screening alone isn’t enough. The federal government should also support a distributed network of biomanufacturing sites, ready to produce countermeasures in a crisis. These facilities should be maintained on standby, stocked with key materials, and able to shift quickly from routine work to emergency response. Redundancy and geographic spread will be critical for avoiding supply chain failures and ensuring resilience.

Modernizing the regulatory infrastructure is no less important. Existing approval processes, particularly within the FDA, are not calibrated for the compressed timelines required during biological emergencies \cite{MCCORMICK2006S63,10.1001/jama.2013.282542}. Regulatory pathways must be adapted to accommodate AI-generated therapeutics by establishing pre-authorized fast-track mechanisms, conditional emergency approvals with post-deployment monitoring, and adaptive clinical trials that integrate real-time safety data. These systems must decisively faster than those designed for conventional drug development. Moreover, regulatory reform must be accompanied by operational flexibility. Pre-positioned decision authorities within public health agencies must be empowered to act swiftly during emergencies, avoiding bureaucratic delays that can be fatal when threats emerge at digital speed.

To ensure the system works in practice and not just in principle, the entire countermeasure pipeline must undergo regular stress testing. Inter-agency simulations—coordinated by the National Security Council, BARDA, ASPR, the Department of Defense, and other relevant bodies—should rehearse the full sequence from biothreat sequence detection to therapeutic deployment. These simulations should include live exercises as well as tabletop models, and must evaluate not only technical performance but also institutional responsiveness. Academic institutions and research hospitals can participate in these simulations by contributing experimental capabilities, validation tools, and scenario modeling.

In parallel, the federal government must create long-term funding mechanisms to sustain this infrastructure. The private sector is unlikely to build and maintain readiness platforms that may be activated infrequently, and commercial incentives are poorly aligned with public health preparedness. Agencies such as NIH, NSF, and DARPA must provide stable support for AI therapeutic design systems, screening platforms, and biomanufacturing capacity, with a mandate for dual-use deployment—available in peacetime for research and ready in crisis for defense. Incentive-compatible public–private partnerships will be crucial. Eligibility for research grants, fast-track regulatory pathways, and procurement contracts should be conditioned on compliance with biosafety, cybersecurity, and auditability standards. Limited liability protections for emergency deployments may be appropriate in high-consequence scenarios, provided transparency and safety are preserved.

In total, the shift from prevention to resilience demands contributions from every level of the scientific ecosystem. Federal agencies must lead on infrastructure, manufacturing, validation, and regulation. Academic institutions must advance the technical foundations, build data infrastructure, and push for therapeutic alignment. The private sector must be incentivized to integrate safety and readiness into commercial innovation. The risks posed by generative biology are real, but so are the opportunities. If we move quickly, the very systems that could accelerate biological threats can also become our most powerful tools for stopping them.

\section{Conclusion}

The rapid emergence of generative AI tools capable of designing novel biological sequences has introduced both new opportunities for biomedical discovery and new vectors for misuse. These models can produce protein candidates with no clear evolutionary antecedents, obscuring their potential function and complicating downstream risk assessment. As our case study illustrates, relying on post hoc filters—such as sequence similarity searches or PPI prediction—to assess the safety of generated sequences is insufficient. Sequence-based heuristics often fail to capture functional novelty, while PPI predictors remain too inaccurate and poorly calibrated to reliably distinguish benign from hazardous outputs.

These limitations expose a broader structural issue: inference-time interventions are fundamentally reactive and speculative. In the absence of reliable, scalable experimental verification, filtering alone cannot guarantee safety. Worse still, an overreliance on these techniques may create a false sense of security, delaying investment in more substantive safeguards.

To address these challenges, we propose a reorientation of biosecurity policy from prevention through prediction toward resilience through response. This includes investment in infrastructure for rapid therapeutic design, high-throughput screening, scalable biomanufacturing, and agile regulatory oversight. The goal is not to eliminate all risk at the point of generation but to ensure that emerging threats can be identified and mitigated in real time.

Generative models will undoubtedly play an increasing role in shaping the future of biology. Ensuring that their benefits are realized without compromising safety requires a realistic appraisal of current technical limitations and a policy framework grounded in robustness rather than prediction alone. As these technologies advance, it is incumbent upon both researchers and policymakers to build systems capable not just of anticipating threats, but of responding swiftly and effectively when they arise.

\section{Data and Model Availability}
The SpatialPPIv2 model used for the analyses in this manuscript can be accessed at this \href{https://github.com/ohuelab/SpatialPPIv2.git}{GitHub repository}. Likewise, the AF3Complex model used can be accessed at this \href{https://github.com/Jfeldman34/AF3Complex.git}{GitHub repository}. The AlphaFold 3 model used is hosted by Google DeepMind on this \href{https://alphafoldserver.com/}{web server}. 
 		
The PDB reference structures, which include both the resolved protein interactions and associated sequences, for the thirteen viral–receptor interaction case studies are as follows. For SARS-CoV-2, the reference structure is 6M0J. For MERS-CoV, the structure is 4L72. The HIV interaction is represented by 1GC1, and the adenovirus interaction is captured by 1F5W. Herpes Simplex Virus interactions are represented by two structures: 1JMA for the HVEM receptor and 3SKU for the Nectin-1 receptor. The Hepatitis B Virus is represented by 8HRX. The Epstein-Barr Virus interaction is described by structure 8SM0. Two distinct measles virus structures are included: 3INB for the Edmonston strain and 4GJT for an alternate strain. Ebola virus is represented by structure 5F1B, Nipah virus by 2VSM, and Chikungunya virus by 6JO8. The sequences for the experimentally validated SARS-CoV-2 mutants, referenced from a previous study \cite{moulana_compensatory_2022}, are publicly available at this \href{https://github.com/desai-lab/compensatory_epistasis_omicron}{GitHub repository}.

\bibliographystyle{unsrtnat}  
\bibliography{Shakhnovich}

\end{document}